**Nucleation and the transition state of the SH3 domain.**

Isaac A. Hubner[†], Katherine A. Edmonds[‡], and Eugene I. Shakhnovich[†*]

[†]Department of Chemistry and Chemical Biology
Harvard University
12 Oxford Street
Cambridge, MA 02138

[‡]Harvard Biophysics Program
Medical School Campus
Building C-2 Room 122
240 Longwood Avenue
Boston, MA 02115

[*]corresponding author
tel: 617-495-4130
fax: 617-384-9228
email: eugene@belok.harvard.edu

**Abstract**

We present a verified computational model of the SH3 domain transition state (TS) ensemble. This model was built for three separate SH3 domains using experimental $\phi$-values as structural constraints in all-atom protein folding simulations. While averaging over all conformations incorrectly considers non-TS conformations as transition states, quantifying structures as pre-TS, TS, and post-TS by measurement of their transmission coefficient ("probability to fold", or $p_{fold}$) allows for rigorous conclusions regarding the structure of the folding nucleus and a full mechanistic analysis of the folding process. Through analysis of the TS, we observe a highly polarized nucleus in which many residues are solvent-exposed. Mechanistic analysis suggests the hydrophobic core forms largely after an early nucleation step. SH3 presents an ideal system for studying the nucleation-condensation mechanism and highlights the synergistic relationship between experiment and simulation in the study of protein folding.

Current understanding of the protein folding process is largely based on the theory of nucleation[1,2] and energy landscape theory[3]. For the majority of small proteins, which serve as model systems for studying protein folding, the process occurs via a two-state mechanism[4]. In analogy to a first order phase transition, the folded and unfolded states may be thought of as different phases (i.e. liquid and gas) and the transition state (TS) as the "critical nucleus" that seeds the transition[5]. Since the kinetics of the transformation between the thermodynamically stable folded and unfolded states is determined by the TS, understanding this state allows for a deeper understanding of the rate and mechanism of folding. The folding process may also be thought of statistically with the formation of a transition state ensemble (TSE) as the rate limiting process in folding[5]. Experiments, mainly based on the protein engineering method of "$\phi$-value analysis"[6], have made significant progress in testing theories of folding. Concurrently, theory and simulation have continued to improve our understanding of the nature of nucleation in the TSE[7].

A prototypical example of the interplay between theory and experiment is the structural interpretation of experimental $\phi$-values through simulation[8,9]. This method for reconstructing a protein's TSE from experimental constraints has proven highly useful in understanding folding, especially as it was refined to test the assumption that a set of structures meeting experimental constraints is, in fact, a TSE[10,11]. A TS is characterized by a transmission coefficient of 1/2. In proteins, this corresponds to a probability of folding, or "$p_{fold}$"[5] equal to one half. Verifying this condition prior to making statements regarding computational models of the TSE is absolutely essential as it has been shown that an ensemble of structures meeting a set of $\phi$-value restraints is not composed entirely of TS conformations[11,12]. Once a model of the TSE is obtained, it is possible to conduct analyses that are currently impossible by experiment, such as exploring the role of residues for which mutation does not cause sufficient destabilization and building an atom-level structural model of the TSE.

SH3 is a widely studied protein that exhibits two-state folding, and is composed of two orthogonally packed three-stranded β-sheets that form a single hydrophobic core[13]. SH3 domains have served as the experimental[13,14] and computational[15-18] model for numerous recent studies of nucleation in protein folding. Although there have been simulations that attempted to reconstruct the TSE using $\phi$-values[17,18], the essential step of testing the assumption that the resulting structures are transition states was not necessarily taken[18]. As it has been shown that not all conformations consistent with experimental $\phi$-values are members of the TSE, it is difficult to accurately interpret these conclusions regarding nucleation and the TSE. In the following study we use the src, fyn, and spectrin SH3 domain and reported experimental $\phi$-values[19] as a model systems to study the nature of nucleation in the protein folding transition state. Importantly, $p_{fold}$ analysis is used to quantify the position of each structure along the folding pathway. We also examine the role and formation of the hydrophobic core and explore the order of folding events in this classic nucleation-condensation mechanism. In SH3, nucleation is early and separate from hydrophobic core formation. The early, polarized TS that we observe is largely solvent-exposed and is formed by a minimal number of contacts.

**Results.**

**A Model of the TSE**. A side-by-side comparison of the native and transition state is presented in Figure 1. The general features of all observed SH3 TSEs include denaturation of the N- and C-termi, turns and loops, and a small amount of secondary structures located in the central β strands. The TSE for the three different SH3 domains exhibit average $C_\alpha$ root mean square deviation (RMSD) ranging from 7.1 to 11.4 Å and radii of gyration ($Rg$) of 12.5 to 14.53Å, which is largely expanded compared to the ~10 Å $Rg$ of the native conformations. The high RSMD is largely a result of unfolding of the N- and C-terminal strands as well as the expansion of the hydrophobic core. The fraction of native contacts ($Q$) ranges from 0.17 in src and spectrin to 0.27 in fyn. By any measure, we observe a TSE in all three domains that occurs early in folding. This implies much less structure than has been observed in streptococcal protein G[11] and ribosomal protein S6[12]. The properties of the pre-TS, TS, and post-TS ensembles for three SH3 domains are summarized in Table 1.

The picture of a TSE that lacks much of the long range interactions found in the native state is consistent with experimental evidence of similarities between the denatured state and folding transition state of the SH3 domain[20]. Our results suggest an early TS, resembling the unfolded ensemble more than the native state, with a small amount of structure (specific nucleus) focused on the central β strands and the majority of the hydrophobic core unformed. Explicit solvent molecular dynamics (MD) studies have also suggested an early TSE with a solvated hydrophobic core[21]. To substantiate and understand the details of the observed diffuse polarized TSE it is essential to identify the specific residues around which the polarized nucleus is organized. Also, it is important to understand the extent to which hydrophobic core formation makes its enthalpic contribution before, during, or after nucleation and, if it is not formed, the extent to which it is exposed to solvent. After understanding the central event of nucleation, we describe the mechanism by which it is formed and how folding proceeds to the native state.

**A distinct network of nucleating contacts.** As expected, in each domain the high $\phi$-value residues that were constrained in simulation generally exhibit a high $Q$ and number of native contacts ($N$). These high $\phi$-value residues, which are experimentally known to play an important role in the TSE, allowed us to effectively sample $p_{fold} = 1/2$ structures. However, they do not uniquely define the TSE. The independently determined and $p_{fold}$ verified TSE of each of the three studied SH3 domains, despite minor differences, exhibit fundamentally similar topology. Figure 2 illustrates the number of contacts made by each residue in the TSE and native states as well as the location of restrained residues, which varied from domain to domain. While most residues make few, if any contacts, a subset of residues confined to a small region of the protein define the TSE. This polarization of structure is characteristic of a classic nucleation mechanism. More important than the number of contacts made by each residue is the information regarding how those contacts are made. In each of the domains, the RT loop, turn/β2, β3, and distal hairpin regions form a network of tertiary interactions. The contact maps of the native and transition states, presented in Figure 3, illustrates this network of contacts and the specific tertiary structural elements that define the native and transition states.

Residues that were restrained in construction of the src TSE due to their high $\phi$-values include E30 and L32 in the diverging turn/β2 regions as well as L44, S47, L48, T50, Q52, T53, Y55, and I56, which span the β3-β4 hairpin[19]. In the fyn domain F4 in β1; F20, E24, and I28 in the RT/turn/β2 region; and T44, G45, I50, and V55 in the

β3/hairpin/β4 region were restrained [22,23]. In spectrin the $\phi$-values of V23, T24, and D29 in the RT/ turn/β2 region; K43 and V44 in β3; and F52, V53, and V58 in β4/β5 were used[24]. Although necessary, the contacts of the restrained residues do not alone define the TSE, as demonstrated through $p_{fold}$ analysis. The fact that even the set of all available restraints does not uniquely define a TSE is also observed in other proteins[11,12]. In SH3, we find that additional restraints do not significantly improve TSE sampling, whereas using fewer restraints results in largely unfolded, low $p_{fold}$, ensembles. Then, by definition, the specific nucleus is composed of the network of contacts formed in each domain by the above residues, including contacts made to residues outside this set in $p_{fold}$ = 1/2 conformations.

From Figure 2, it is clear that a number of unconstrained residues also play a role in the TSE. There is a small amount of structure in the RT loop centered on residues near the diverging turn. L24 (numbering from the src domain) is a highly conserved position in the SH3 fold family[25] and has been shown experimentally to be at least partially involved in the TSE[26] despite its relatively low $\phi$-value and also appears to play an important role in the fyn and spectrin domains. The n-src loop has some native structure around Q33 and I34 near β2, as well as D41, which leads into β3. None of these residues have reported $\phi$-values. The same effect is seen in the fyn and spectrin domain, where the n-src loop region is unstructured. Simulation provides key insight into the structure of the n-src loop in the TSE, since most experimental data on this region is difficult to interpret[19]. In the β3/distal hairpin region, W43 and A45, both highly conserved in SH3, make a number of important contacts in the TSE. W43 has no $\phi$-value and A45 has a difficult to interpret $\phi$ = 1.20. Residue G51, which is also highly conserved across different SH3 domains[27], exhibits high $Q$ but very low values of $N$. Its important role is also suggested by an experimental $\phi = 1.06$[19] as well as other recent, independent mutagenic experiments[27].

Through the preceding analysis, it is clear that the structure of the TSE, common to all domains, is formed by a small number of residues in the β2-β3 hairpin making non-local contacts to the RT loop/diverging turn and distal β-hairpin regions. The nucleus residues common to all three domains include L24, F26, E30, L32, W43, L44, A45, H46, and G51 (numbering corresponds to the src domain, see Supplemental Figure 1 for an alternative presentation of the data). This agrees with experimental results indicating that the second, third, and to a lesser extent the fourth β strands are the most ordered regions of the TSE[19]; however, we are able to go beyond this qualitative description to understand how the contacts are being made (Figure 3). The importance of this network of tertiary contacts in the SH3 TSE has also been observed in simplified model simulations[28]. It is interesting to note that the folding nucleus is not simply composed of large hydrophobic residues as one might naively assume. In fact many such residues belong to "unfolded" regions in the TSE. The network of contacts formed by the specific nucleus described above is the necessary and sufficient condition for reaching the TSE. It is noteworthy that simply averaging over all conformations consistent with experimental restraints yields a misleading picture of a TSE that is largely defined as a distorted/partially formed native state topology, a claim that has been made previously in SH3[18]. Accurate characterization of the true TSE is only possible through the implementation of $p_{fold}$ analysis, which allows for the discrimination and separate analysis of pre-TS, TS, and post-TS conformations.

**Solvent accessibility in the TSE**. Many experimental[29,30] and theoretical[21,31] studies have used SH3 domains to explore and understand the role of solvent in protein folding. Since the Go model does not explicitly represent solvent, we examine solvent accessibility in the TSE rather than the mechanistic role of individual water molecules. Figure 4 summarizes the change in solvent accessibility for each residue between the native and transition states for three SH3 domains. The N- and C-terminal strands are both very solvent exposed, indicating their largely denatured conformation in the TSE. This may also be inferred from Figure 2, which shows how few contacts these residues make. More striking than the solvent exposure of these residues, is the fact that the most exposed residues in the first β strand and beginning of the RT-loop, and the residues from the end of β4 through the $3_{10}$ helix and β5 are largely non-polar and hydrophobic. The same trend of exposed hydrophobic residues is observed in the n-src loop and beginning of the β3 strand. This observation is consistent across three different SH3 domains.

As described above, full formation of the hydrophobic core is not necessary for reaching the TSE. The ordering of water molecules around these hydrophobic residues creates an entropic penalty that is relieved upon desolvation. Thus, desolvation in the polymer compaction subsequent to the transition barrier, may mitigate the entropic cost of polymer collapse. This, coupled with the energy of the eventual hydrophobic contact formation, could help ensure that subsequent folding is rapid and highly energetically favorable. It has been suggested that these residues may be involved in non-native or non-specific hydrophobic interactions[29]. Since a strict Go potential treats non-native contacts as destabilizing, we were concerned that such a model could skew any test of this hypothesis. As such, we also performed $p_{fold}$ analysis with an alternative potential that does not count all non-native interactions as repulsive and which includes attractive non-native interaction. This generalization of the model did not alter the basic result; both with and without non-native interactions, our simulations suggest that such non-specific hydrophobic stabilization is necessary neither for nucleation nor rapid folding. Although negative experimental $\phi$-values are sometimes interpreted as indicating non-native interactions or parallel folding pathways[32], nearly all residues in the hydrophobic core have positive $\phi$-values. Our model also suggests that the energetic contribution of hydrophobic core formation comes only after the transition barrier.

The other major, though less pronounced, region of increased solvent accessibility in the TSE spans the RT-loop into the diverging turn and β2 strand. This group of residues is different from the one discussed above in that they are predominantly polar or charged. If hydrophilic residues are exposed in the TSE, then solvation of these residues may provide non-specific stabilization, accelerating folding. This interpretation is consistent with the early barrier observed independently in earlier, explicit solvent MD models that seek to determine the TSE from unfolding[21]. Other simple models have also suggested the important role of solvent in SH3 folding[33]. From our simulations, it appears that a fully formed hydrophobic core is not necessary for traversing the transition barrier in SH3. In fact, the solvent exposure of these core residues may play an important role in promoting folding past the TSE by affecting the energetic balance between desolvation, configurational entropy loss, and hydrophobic contact formation.

**Mechanistic insights**. It was previously demonstrated that the order of events in protein folding may be reliably inferred from examination of structures that precede and follow the TSE along the folding pathway[12]. Figure 3 (right column) shows the contacts present in the pre- (but not post-) TSE and post- (but not pre-) TSE structures. Double mutant experiments show that the distal hairpin is mostly formed in the TS[14] and there is some indication that is begins to form prior to the TSE in all three domains. The minute amount of tertiary structure present before the transition state occurs is greatly fluctuating and varies between the different domains, which is consistent with an early TSE. Conversely, there are a large number of structural features common to the TSE of each domain. Despite minor differences, the common behavior we observe between all studied SH3 domains is the formations of the turn-RT loop, β1-β2, β2-β3, and β3-β4 motifs immediately after the TSE. Contacts between the β2 and β3 strands begin forming around the transition region to various degrees, but form largely after nucleation. Cross linking experiments have also suggested that the N- and C-termini and RT loop form after the rate limiting step[14]. This behavior is also observed in our simulation with a slight exception in spectrin, which exhibits contacts between the $3_{10}$ helix region and β1 in the TSE. The n-src loop is anchored by β2 and minimally solvent-exposed in the TSE, but appears to adopt a native conformation late in SH3 folding.

**A classic nucleation-condensation mechanism**. Through simulation, we observe that SH3 folding proceeds via a classic nucleation-condensation mechanism. A small number of non-local contacts form the specific nucleus. Though some previous studies focused on the β2-β3-β4 sheet as the defining characteristic of the TSE[17], we observe that tertiary contacts involving the RT-loop, turn/β2, β3 region, and the distal hairpin characterize $p_{fold}$ = 1/2 conformations. The importance of tertiary contacts in SH3 domain transition states has previously been observed in simplified model simulations[16]. However, some of the residues they identify such as A12 ($\phi = 0.05$) and G54 ($\phi = -0.08$) do not appear to play a major role in the nucleation. This highlights the advantages inherent in the increased resolution of an all-atom model over a more coarse grained Cα/Cβ protein representation. One computational study suggested that the β2-β3-β4 sheet is fully hydrogen bonded with little other structure in the TSE and that folding occurs through a diffusion-collision like model of sheet formation, followed by a separate sandwiching event and expulsion of water from the core[21]. In contrast, we observe the RT loop, turn-β2 motif, β3 strand, and the distal β-hairpin involved in a network of non-local tertiary contacts forming the site of nucleation.

Our results are consistent across experiments on several different SH3 domains and with independent simulations that suggest that the distal hairpin and loop are ordered in the SH3 TSE, while the rest of the protein is largely disordered[15]. Experiments in α-spectin SH3 have also shown that loop stiffness is important for restricting structural heterogeneity in the TSE[34]. The nucleus is distinct from the hydrophobic core, which is mainly unformed and solvent-exposed in the TSE[34]. The nucleus occurs in an early, polarized TSE that requires relatively few native contacts. The robustness of folding to environmental changes is evidenced in experiments that show that nucleation is unaffected even by large changes in pH[24]. It has also been observed in the D48G mutant of α-spectrin domain that the protein melts over a 20° range[35]. One rationalization for this observation, may be that since hydrophobic core formation it separate from nucleation, the non-cooperative "melting" of these contacts broadens the transition. The highly cooperative behavior of the SH3 domain[13], and the observed consistency of folding mechanism despite drastic changes in experimental conditions, may be attributed to the early, polarized TSE formation by a minimal number of contacts. These data provide an atomic-level description of a conformationally restricted folding transition state in which a small region (the RT loop, β3, β3-β4 hairpin and its contacts to β2 and the diverging turn, a network formed largely by L24, F26, E30, L32, W43, L44, A45, H46, and G51) is the center of nucleation and in which the rest of the protein, including residues involved in the hydrophobic core, is largely disordered[24]. We observe that nucleation comes before, and is separate from, hydrophobic collapse. Simple models have also suggested that formation of the hydrophobic core may occur separately, and within a distinct timescale, from structural collapse to a "near native intermediate"[33]. The early nucleation we observe is the rate-limiting step and is distinct from the late native-like structures found by others; what they refer to is not nucleation, but a separate desolvation step[30,31]. MD models with explicit solvent also suggest that core formation and packing come after β-sheet formation[21]. This separation of timescales may also be explained as resulting from the kinetically distinct phase of side chain relaxation that follows nucleation and collapse[36].

**Discussion.**

We have presented a $p_{fold}$ verified, all-atom model of the TSE in three different SH3 domains. This allows for rigorous conclusions regarding the nature and mechanism of nucleation. In constructing the TSE, residues were restrained, but the way in which the minimal set of restraints was met (contacts between residues) was not predetermined. Importantly, the true TSE was identified through (unconstrained) $p_{fold}$ analysis. It is noteworthy that the set of available $\phi$-values for SH3, and other proteins, is insufficient to uniquely specify the TSE. One recent study reports an unverified TSE constructed from $\phi$-values in which ensembles produced under different unfolding conditions have virtually non-overlapping RMSD/$Rg$ distributions[18]. Another study acknowledged the presence of "false positive TSE structures", but relied on manual selection of structures such that their model TSE was in agreement with presumed general features such as a formed β2-β3-β4 sheet[17]. In the absence of a quantitative measure, it is impossible to reliably recognize the transition state. However, with a suitable proxy for the transition coordinate such as $p_{fold}$[5], it is possible to identify the distinct structural characteristics of the TSE and other stages along the folding pathway.

From these data we conclude that SH3 folds via a classic nucleation-condensation mechanism with a highly polarized nucleus that forms early along the reaction pathway, which helps explain the experimentally observed highly cooperative folding of SH3[13]. In this early TSE many residues (both hydrophobic and hydrophilic) are solvent-exposed. Hydrophobic collapse is separate from, and subsequent to, nucleation. In this sense SH3 is the paradigm for the nature and implication of a pure nucleation-condensation protein folding mechanism. While experimental $\phi$-values are invaluable in helping to identify residues of importance in folding, simulation allows for deeper interpretation by determining the identity of residues and network of contacts that form the specific nucleus. In this manner, kinetic $\phi$-values are structurally interpreted as a completely interacting set of data in the context of a 3D protein structure. We believe that similar studies, in which computer simulations are utilized to interpret experimental data, will play an increasingly important role in understanding protein folding.

**Model and Methods**

**Constructing and verifying the TSE**. Simulations were initiated using coordinates from the crystal structures of the src-SH3 of tyrosine kinase[37] (1FMK), fyn-SH3[22,23] (1FYN), and α-spectrin SH3[24] domains (1BK2).. The simulation was propagated by Monte Carlo dynamics using a move set that included rotations of backbone ($\phi$-$\psi$) angles and sidechain ($\chi$) torsion angles, while maintaining planar peptide bonds. All non-hydrogen atoms are represented as impenetrable hard spheres[38,39] and excluded volume is continually enforced. Energy was provided by a square-well Go potential since these models faithfully represent many of the important features of the protein folding landscape in a computationally efficient manner[40]. As in experiment[13], our model of SH3 exhibits clear two-state unfolding without intermediates. The thermodynamics and kinetics of this model (potential, move set, and protein representation) have been extensively studied in crambin[38], protein G[11,39], CI2[10], and S6[12], successfully folding these proteins from random coil to <1Å $C_\alpha$ RMS.

The SH3 folding mechanism is largely determined by native state topology, with nonnative interactions playing a relatively small role[13]; thus the native-centric Go model is especially well suited for its study. Use of the Go model is also motivated by rigorous theoretical and experimental studies showing that the TS is robust with respect to the selection of sequences folding into a particular structure and potentials used to design and fold sequences[41-43]. Go models have been applied to the prediction of folding rates and the interpretation and prediction of $\phi$-values[44-46]. Since there is no general physics-based (CHARMM-like) potential for fully folding α/β from random coil, the topologically-based and computationally efficient Go potential is currently the ideal candidate[40] for protein folding studies such as these. It was recently demonstrated that Go models compare favorably with empirical potentials, especially in representing the interaction energies of individual residues and in interpreting $\phi$-values[47]. A putative ensemble of TS conformations was constructed through restrained unfolding simulations, and $p_{fold}$ determined for each member in order to verify if a given structure is a TS, as earlier described[11,12]. Experimental $\phi$-values ($\phi^{exp}$) were interpreted as the fraction of native contacts made by a given residue in the TSE ($\phi^{sim}$):

$$\phi_i^{sim} = \frac{N_i^{TSE}}{N_i^{Native}} \quad (1)$$

and implemented as structural restraints in unfolding simulations by use of a simple harmonic potential for restrained residues:

$$E^{Total} = E^{Go} + E^{\phi}$$
$$= E^{Go} + \Lambda \cdot \sum_i^N (\phi_i^{sim} - \phi_i^{exp})^2 \quad (2)$$

It is intuitively and desirable to use the minimal number of restraints necessary to construct the putative TSE (none are used in $p_{fold}$ simulation) and to choose those restraints such that most straightforwardly interpretable. To this end, we exclude $\phi > 1$ and $\phi < 0$ (which may indicate native contacts, non-native contacts, or alternative folding pathways). These experimental values are without doubt meaningful, however ambiguity in their interpretation and the fact that they are not necessary to construct the TSE argue convincingly for their exclusion. Previously published experimental $\phi$-values[19,22-24] were used as a source of restraints. For src-SH3, we chose the 10 highest $\phi$-values after excluding all $\phi > 1$ and V35, which is in a cluster of six residues for which mutation yielded unclear results. Both fyn and spectrin, exhibited 8 classical, unambiguous values.

In the highly multi-dimensional energy landscape of protein folding, it is difficult to define a reaction coordinate or progress variable by which to gauge the evolution of the folding reaction. Several practical solutions to this problem have been proposed[48,49]. We choose to apply $p_{fold}$ analysis which, with firm grounding in theory and simulation[48], defines a measurable universal property of chemical systems that may serve as a progress variable. By definition, a system that exhibits a $p_{fold} = 1/2$ is a TS. The adaptation of $p_{fold}$ analysis to protein models systems has been detailed previously[38]. We initiate 100 independent restrained unfolding simulations were initiated from the src-SH3 native coordinates. 50 unfolding runs each were conducted for the fyn and spectrin domains. Unfolding trajectories were run at T = 2.5 ≈ $T_f$ (4 for spectrin). Each independent unfolding simulation was followed by 20 unrestrained refolding runs to determine $p_{fold}$. Refolding simulations were run under stabilizing conditions at T = 1, to observe rapid refolding of structures that have passed the transition barrier. The criteria for commitment to folding were derived from the distribution of values of the equilibrium ensemble at the refolding T (E < -425, N > 425, DRMS < 2.5 Å). The TSE includes structures with $0.4 \leq p_{fold} \leq 0.6$. Structures with lower $p_{fold}$ are considered "pre-TS" and with higher $p_{fold}$ "post-TS". The properties of these ensembles are summarized in Table 1. In the TSE we observe a correlation of R = 0.82 between our computed and experimental $\phi$-values.

**How does the potential influence $p_{fold}$?** The fact that the Go potential is constructed from knowledge of a protein's native state and that each atom is unique may lead to the unphysical situations where, for example, one (native) phenylalanine – phenylalanine interaction is attractive, while another (non-native) is not. So while Go serves as an extremely useful model that abstracts any of the fundamental theories relating to protein folding[40], alone it cannot meaningfully address non-native interactions.

If upon passing the transition barrier, descent into the native basin of attraction is rapid and energetically downhill, then one may ask the question: how does the Go model, with its idealized, smooth funnel-like representation of the energetic landscape, effect the modeling of this process? In order to explore the manner in which attractive and repulsive non-native interactions might affect our $p_{fold}$ calculations, we hybridize the Go (which ensures folding to the native state) with the μ potential (first described in the folding of the streptococcal protein A[50]), which treats all interactions as physically attractive or repulsive. A similarly structured hybrid potential (with 90% Go) was recently used to study the role of non-native interactions in RNA folding[51].

The μ potential is calculated according to:

$$E_{AB} = \frac{-\mu N_{AB} + (1-\mu)\tilde{N}_{AB}}{\mu N_{AB} + (1-\mu)\tilde{N}_{AB}} \quad (3)$$

where $E_{AB}$ is contact interaction energy between atom types $A$ and $B$, $N_{AB}$ is the number of $AB$ pairs found in contact and $\tilde{N}_{AB}$ is the number of $AB$ pairs in the database that are not in contact. For this work, μ was chosen (0.9979) such that $N_{AB}$ ranges from –1 to +1 and has an average value of 0. Contact statistics were collected from a database of 103 non-homologous proteins with less than 25% sequence homology, longer than 50, and shorter than 200 residues[52]. Atom types (Supplemental Table 1) were chosen such that each unique atom (excluding hydrogen) of each residue sidechain is a unique type. Backbone atoms are typed without regard to residue identity.

Here, we repeat the $p_{fold}$ calculations for the 100 conformations created by the restrained unfolding simulations of src-SH3 domain with a hybrid of $E_{AB}$ (Equation 3) and 75% $E_{Go}$. Additionally, we set up the Go potential such that native interactions have a -1 attraction and non-native interaction have no interaction (as opposed to the +1 repulsion of the strict Go potential). Thus, we remove the funnel-like bias towards the native state and introduce meaningful attractive and repulsive non-native interactions. In order to maintain uniformity in the interpretation of the results, the same criteria we used to define a folded structure. Despite the significantly different (more frustrated) energy landscape, the potential finds the native basin quickly in high $p_{fold}$ conformations, and unfolds in low $p_{fold}$ conformations. In fact the correlation for 100 $p_{fold}$ values for the two sets is R = 0.85, which strongly suggests that $p_{fold}$ is not sensitive to the details of the potential.

**Acknowledgements**

We thank Dr. Brian N. Dominy for assistance with CHARMM, and Eric J. Deeds for his comments on the manuscript. This work was supported by an HHMI pre-doctoral fellowship (IAH).

**References**

1. Mirny, L. & Shakhnovich, E. Protein folding theory: from lattice to all-atom models. *Annu Rev Biophys Biomol Struct* **30**, 361-96 (2001).
2. Galzitskaya, O.V., Ivankov, D.N. & Finkelstein, A.V. Folding nuclei in proteins. *FEBS Lett* **489**, 113-8 (2001).
3. Onuchic, J.N. & Wolynes, P.G. Theory of protein folding. *Curr Opin Struct Biol* **14**, 70-5 (2004).
4. Jackson, S.E. How do small single-domain proteins fold? *Fold Des* **3**, R81-91 (1998).
5. Pande, V.S., Grosberg, A., Tanaka, T. & Rokhsar, D.S. Pathways for protein folding: is a new view needed? *Curr Opin Struct Biol* **8**, 68-79 (1998).
6. Matouschek, A., Kellis, J.T., Jr., Serrano, L. & Fersht, A.R. Mapping the transition state and pathway of protein folding by protein engineering. *Nature* **340**, 122-6 (1989).
7. Pande, V.S. Meeting halfway on the bridge between protein folding theory and experiment. *Proc Natl Acad Sci U S A* **100**, 3555-6 (2003).
8. Onuchic, J.N., Socci, N.D., Luthey-Schulten, Z. & Wolynes, P.G. Protein folding funnels: the nature of the transition state ensemble. *Fold Des* **1**, 441-50 (1996).
9. Vendruscolo, M., Paci, E., Dobson, C.M. & Karplus, M. Three key residues form a critical contact network in a protein folding transition state. *Nature* **409**, 641-5 (2001).
10. Li, L. & Shakhnovich, E.I. Constructing, verifying, and dissecting the folding transition state of chymotrypsin inhibitor 2 with all-atom simulations. *Proc Natl Acad Sci U S A* **98**, 13014-8 (2001).
11. Hubner, I.A., Shimada, J. & Shakhnovich, E.I. Commitment and nucleation in the protein G transition state. *J Mol Biol* **336**, 745-61 (2004).
12. Hubner, I.A., Oliveberg, M. & Shakhnovich, E.I. Simulation, experiment, and evolution: Understanding nucleation in protein S6 folding. *Proc Natl Acad Sci U S A* (2004).
13. Grantcharova, V.P. & Baker, D. Folding dynamics of the src SH3 domain. *Biochemistry* **36**, 15685-92 (1997).
14. Grantcharova, V.P., Riddle, D.S. & Baker, D. Long-range order in the src SH3 folding transition state. *Proc Natl Acad Sci U S A* **97**, 7084-9 (2000).
15. Klimov, D.K. & Thirumalai, D. Stiffness of the distal loop restricts the structural heterogeneity of the transition state ensemble in SH3 domains. *J Mol Biol* **317**, 721-37 (2002).
16. Borreguero, J.M., Dokholyan, N.V., Buldyrev, S.V., Shakhnovich, E.I. & Stanley, H.E. Thermodynamics and folding kinetics analysis of the SH3 domain form discrete molecular dynamics. *J Mol Biol* **318**, 863-76 (2002).
17. Settanni, G., Gsponer, J. & Caflisch, A. Formation of the folding nucleus of an SH3 domain investigated by loosely coupled molecular dynamics simulations. *Biophys J* **86**, 1691-701 (2004).

18. Lindorff-Larsen, K., Vendruscolo, M., Paci, E. & Dobson, C.M. Transition states for protein folding have native topologies despite high structural variability. *Nat Struct Mol Biol* **11**, 443-9 (2004).
19. Riddle, D.S. et al. Experiment and theory highlight role of native state topology in SH3 folding. *Nat Struct Biol* **6**, 1016-24 (1999).
20. Kortemme, T., Kelly, M.J., Kay, L.E., Forman-Kay, J. & Serrano, L. Similarities between the spectrin SH3 domain denatured state and its folding transition state. *J Mol Biol* **297**, 1217-29 (2000).
21. Guo, W., Lampoudi, S. & Shea, J.E. Posttransition state desolvation of the hydrophobic core of the src-SH3 protein domain. *Biophys J* **85**, 61-9 (2003).
22. Northey, J.G., Maxwell, K.L. & Davidson, A.R. Protein folding kinetics beyond the phi value: using multiple amino acid substitutions to investigate the structure of the SH3 domain folding transition state. *J Mol Biol* **320**, 389-402 (2002).
23. Northey, J.G., Di Nardo, A.A. & Davidson, A.R. Hydrophobic core packing in the SH3 domain folding transition state. *Nat Struct Biol* **9**, 126-30 (2002).
24. Martinez, J.C. & Serrano, L. The folding transition state between SH3 domains is conformationally restricted and evolutionarily conserved. *Nat Struct Biol* **6**, 1010-6 (1999).
25. Larson, S.M. & Davidson, A.R. The identification of conserved interactions within the SH3 domain by alignment of sequences and structures. *Protein Sci* **9**, 2170-80 (2000).
26. Grantcharova, V.P., Riddle, D.S., Santiago, J.V. & Baker, D. Important role of hydrogen bonds in the structurally polarized transition state for folding of the src SH3 domain. *Nat Struct Biol* **5**, 714-20 (1998).
27. Di Nardo, A.A. et al. Dramatic acceleration of protein folding by stabilization of a nonnative backbone conformation. *Proc Natl Acad Sci U S A* **101**, 7954-9 (2004).
28. Ding, F., Dokholyan, N.V., Buldyrev, S.V., Stanley, H.E. & Shakhnovich, E.I. Direct molecular dynamics observation of protein folding transition state ensemble. *Biophys J* **83**, 3525-32 (2002).
29. Viguera, A.R., Vega, C. & Serrano, L. Unspecific hydrophobic stabilization of folding transition states. *Proc Natl Acad Sci U S A* **99**, 5349-54 (2002).
30. Fernandez-Escamilla, A.M. et al. Solvation in protein folding analysis: combination of theoretical and experimental approaches. *Proc Natl Acad Sci U S A* **101**, 2834-9 (2004).
31. Shea, J.E., Onuchic, J.N. & Brooks, C.L., 3rd. Probing the folding free energy landscape of the Src-SH3 protein domain. *Proc Natl Acad Sci U S A* **99**, 16064-8 (2002).
32. Ozkan, S.B., Bahar, I. & Dill, K.A. Transition states and the meaning of Phi-values in protein folding kinetics. *Nat Struct Biol* **8**, 765-9 (2001).
33. Cheung, M.S., Garcia, A.E. & Onuchic, J.N. Protein folding mediated by solvation: water expulsion and formation of the hydrophobic core occur after the structural collapse. *Proc Natl Acad Sci U S A* **99**, 685-90 (2002).
34. Spagnolo, L., Ventura, S. & Serrano, L. Folding specificity induced by loop stiffness. *Protein Sci* **12**, 1473-82 (2003).

35. Martinez, J.C., Pisabarro, M.T. & Serrano, L. Obligatory steps in protein folding and the conformational diversity of the transition state. *Nat Struct Biol* **5**, 721-9 (1998).
36. Kussell, E., Shimada, J. & Shakhnovich, E.I. Side-chain dynamics and protein folding. *Proteins* **52**, 303-21 (2003).
37. Xu, W., Harrison, S.C. & Eck, M.J. Three-dimensional structure of the tyrosine kinase c-Src. *Nature* **385**, 595-602 (1997).
38. Shimada, J., Kussell, E.L. & Shakhnovich, E.I. The folding thermodynamics and kinetics of crambin using an all-atom Monte Carlo simulation. *J Mol Biol* **308**, 79-95 (2001).
39. Shimada, J. & Shakhnovich, E.I. The ensemble folding kinetics of protein G from an all-atom Monte Carlo simulation. *Proc Natl Acad Sci U S A* **99**, 11175-80 (2002).
40. Takada, S. Go-ing for the prediction of protein folding mechanisms. *Proc Natl Acad Sci U S A* **96**, 11698-700 (1999).
41. Abkevich, V.I., Gutin, A.M. & Shakhnovich, E.I. Specific nucleus as the transition state for protein folding: evidence from the lattice model. *Biochemistry* **33**, 10026-36 (1994).
42. Shakhnovich, E., Abkevich, V. & Ptitsyn, O. Conserved residues and the mechanism of protein folding. *Nature* **379**, 96-8 (1996).
43. Chiti, F. et al. Mutational analysis of acylphosphatase suggests the importance of topology and contact order in protein folding. *Nat Struct Biol* **6**, 1005-9 (1999).
44. Galzitskaya, O.V. & Finkelstein, A.V. A theoretical search for folding/unfolding nuclei in three-dimensional protein structures. *Proc Natl Acad Sci U S A* **96**, 11299-304 (1999).
45. Alm, E. & Baker, D. Prediction of protein-folding mechanisms from free-energy landscapes derived from native structures. *Proc Natl Acad Sci U S A* **96**, 11305-10 (1999).
46. Munoz, V. & Eaton, W.A. A simple model for calculating the kinetics of protein folding from three-dimensional structures. *Proc Natl Acad Sci U S A* **96**, 11311-6 (1999).
47. Paci, E., Vendruscolo, M. & Karplus, M. Validity of Go models: comparison with a solvent-shielded empirical energy decomposition. *Biophys J* **83**, 3032-8 (2002).
48. Du, R., Pande, V.S., Grosberg, A., Tanaka, T. & Shakhnovich, E. On the transition coordinate for protein folding. *J Chem Phys* **108**, 334-350 (1998).
49. Klimov, D.K. & Thirumalai, D. Multiple protein folding nuclei and the transition state ensemble in two-state proteins. *Proteins* **43**, 465-75 (2001).
50. Kussell, E., Shimada, J. & Shakhnovich, E.I. A structure-based method for derivation of all-atom potentials for protein folding. *Proc Natl Acad Sci U S A* **99**, 5343-8 (2002).
51. Sorin, E.J. et al. Does native state topology determine the RNA folding mechanism? *J Mol Biol* **337**, 789-97 (2004).
52. Mirny, L.A. & Shakhnovich, E.I. How to derive a protein folding potential? A new approach to an old problem. *J Mol Biol* **264**, 1164-79 (1996).

Figure 1. Model of src-SH3 native (a) and TSE (b) structures. Regions corresponding to the different β strands are colored from N (blue) to C (red) termini. Loop, turn, hairpin, and helix regions are labeled on the native structure. The TSE is represented by five superimposed conformations.

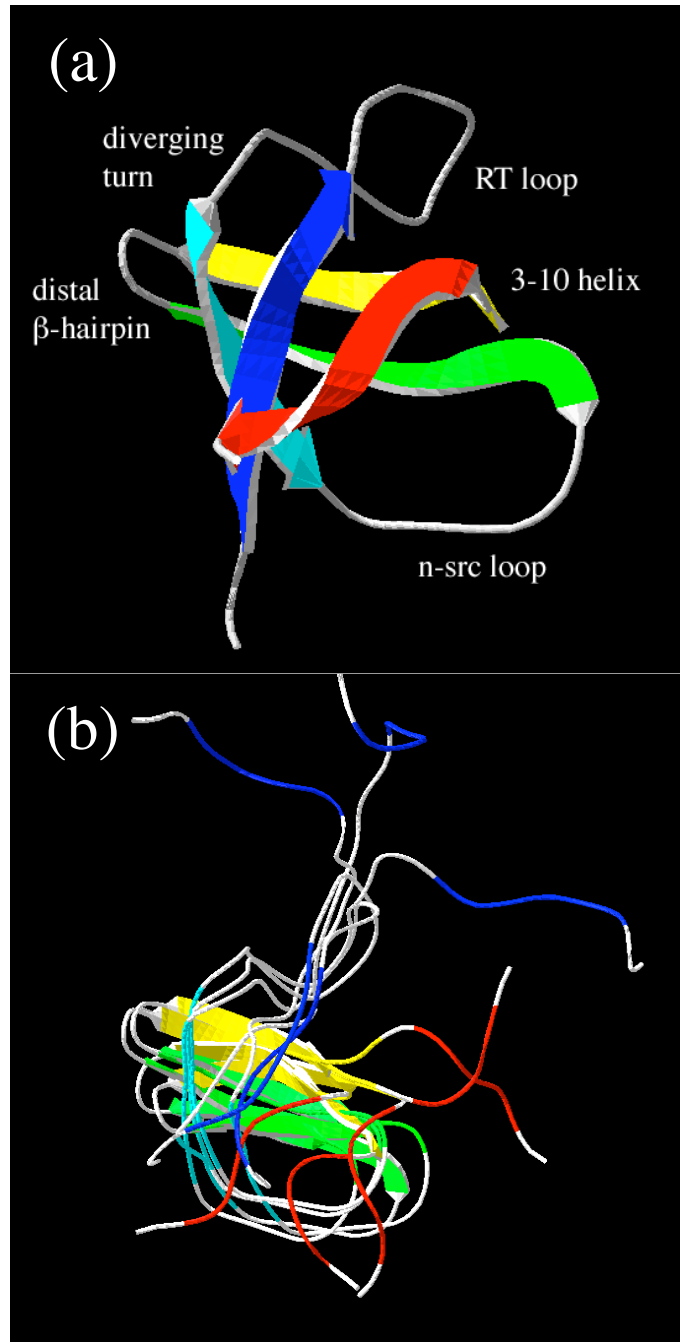

Figure 2. The total number ($N$) of contacts made by each residue in the native state and TSE for the (a) src, (b) fyn, and (c) spectrin SH3 domains. The native state values are plotted in black and those for the TSE are in red. Residue numbers correspond to the respective experimental reference. Structural elements and their residue ranges are labeled at the top and restrained residues are circled.

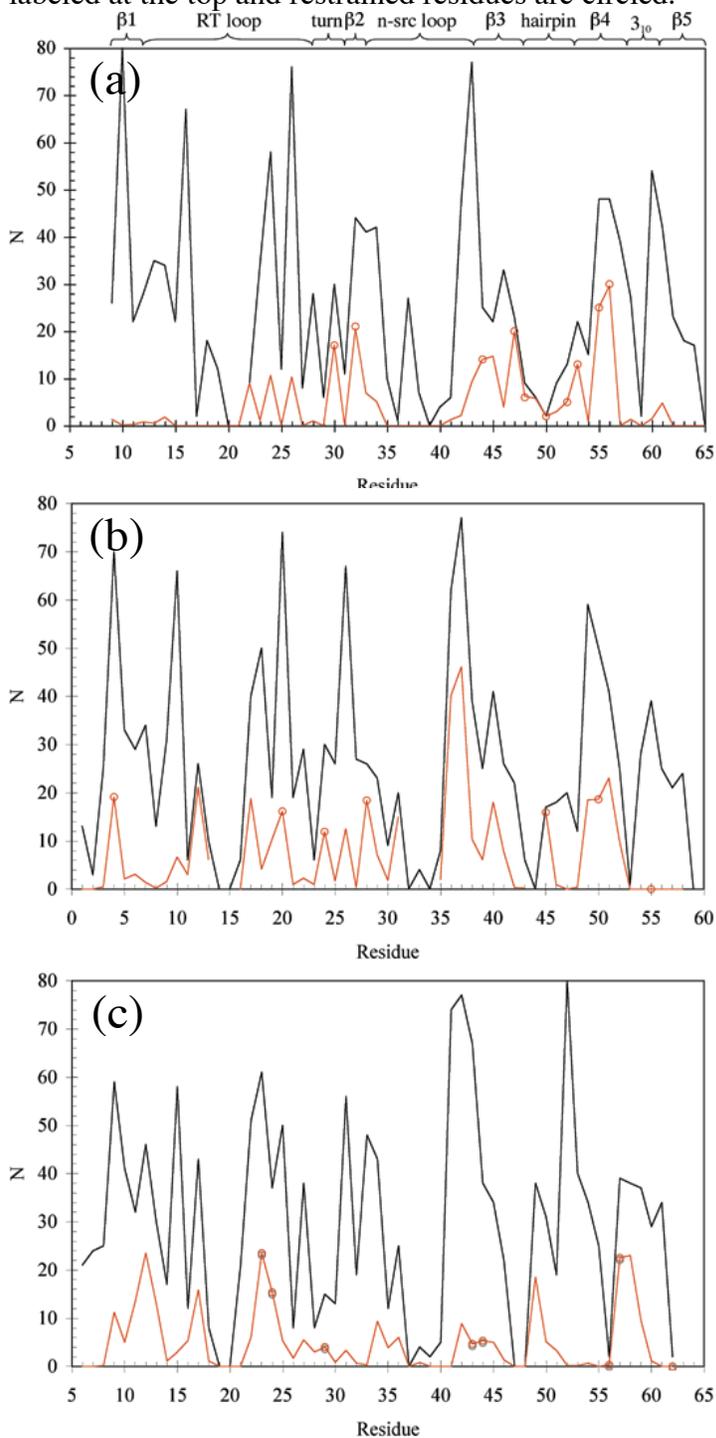

Figure 3. Contact maps comparing the native (upper left) and TSE (lower right) conformations and the pre-TS (upper left) and post-TS (lower right) ensembles for (a) src, (b) fyn, and (c) spectrin. Tertiary structural elements are circled and noted in red. The location of the different β sheets are noted by arrows along the axis, which also indicate the corresponding residue numbers.

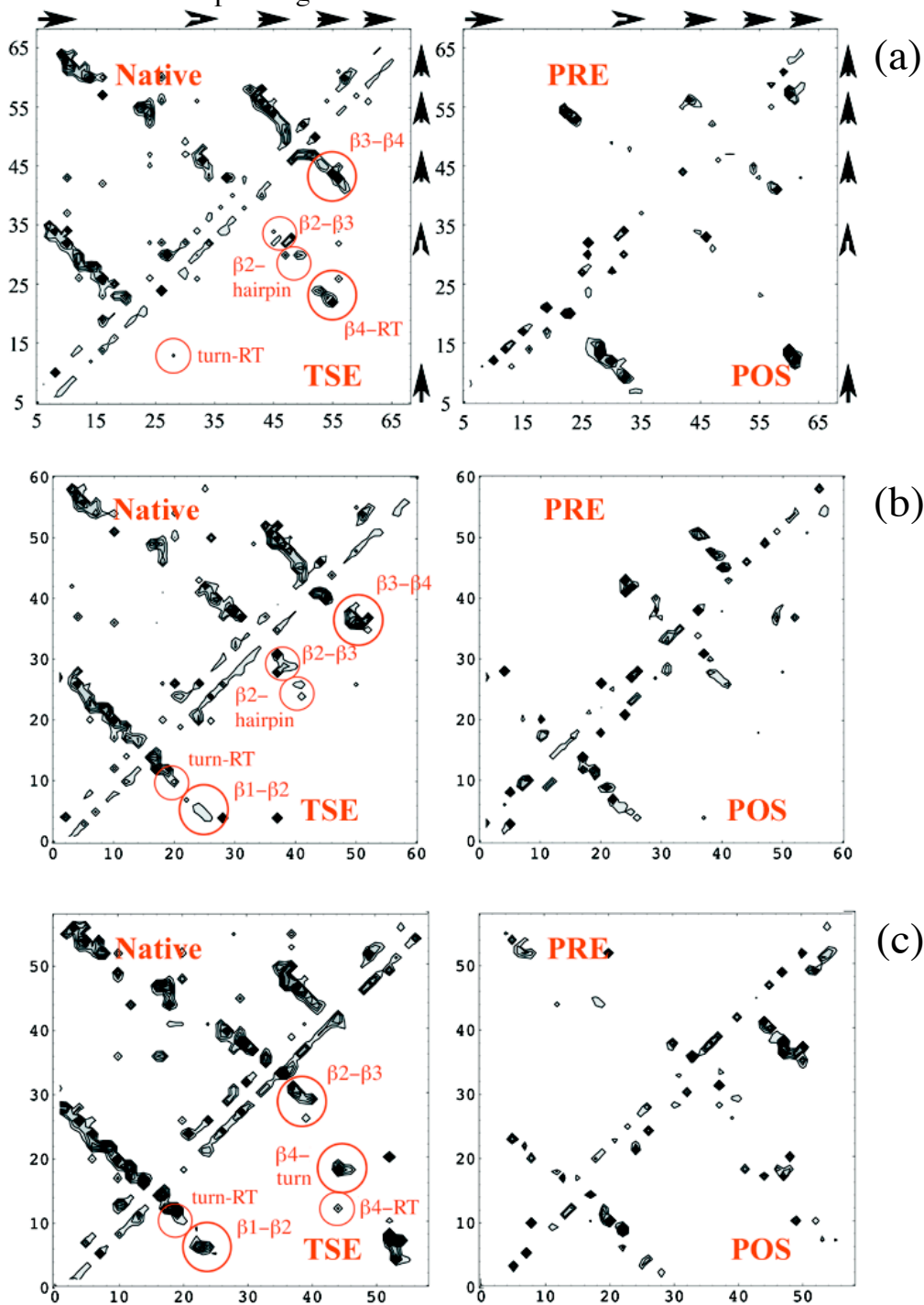

Figure 4. Difference in solvent accessible surface area (SASA) in Å$^2$ between transition and native states for the (a) src, (b) fyn, and (c) spectrin SH3 domains. Non-polar and hydrophobic residues are colored red. Polar and charged residues are colored blue.

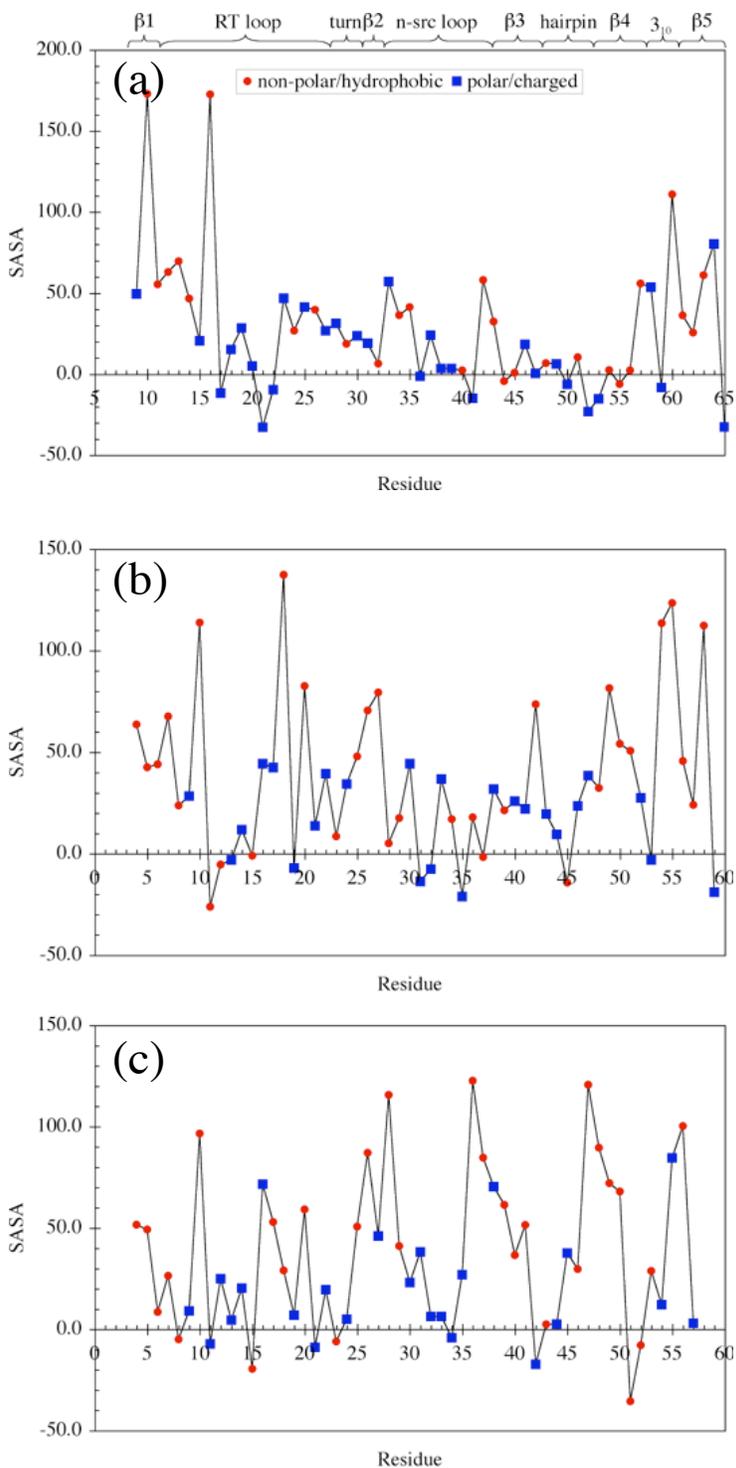

Table 1. Ensemble properties fot the three SH3 domains.

| domain | ensemble | $C_\alpha$ RMSD | $Rg$ | N(native) | N(non-native) |
|---|---|---|---|---|---|
| src | Native | - | 10.14 | 725 | - |
| | Pre-TS | 13.6 ± 3.6 | 15.6 ± 2.1 | 119 ± 4 | 70 ± 13 |
| | TS | 11.4 ± 4.9 | 14.5 ± 2.5 | 124 ± 8 | 70 ± 14 |
| | Post-TS | 4.1 ± 1.1 | 11.2 ± 0.4 | 147 ± 14 | 91 ± 13 |
| fyn | Native | - | 9.97 | 760 | - |
| | Pre-TS | 9.6 ± 1.9 | 13.7 ± 0.8 | 191 ± 32 | 62 ± 15 |
| | TS | 8.7 ± 1.9 | 13.3 ± 0.7 | 203 ± 25 | 60 ± 14 |
| | Post-TS | 7.2 ± 1.5 | 12.9 ± 0.8 | 213 ± 27 | 60 ± 13 |
| spectrin | Native | - | 9.62 | 845 | - |
| | Pre-TS | 8.2 ± 1.2 | 13.0 ± 0.51 | 152 ± 21 | 80 ± 15 |
| | TS | 7.1 ± 0.9 | 12.5 ± 0.49 | 146 ± 31 | 70 ± 14 |
| | Post-TS | 5.76 ± 1.3 | 12.0 ± 0.65 | 189 ± 55 | 88 ± 20 |

Supplemental Figure 1. Alternative presentation of TSE residue contact data for three SH3 domains. All sequences are plotted as aligned to the src domain.

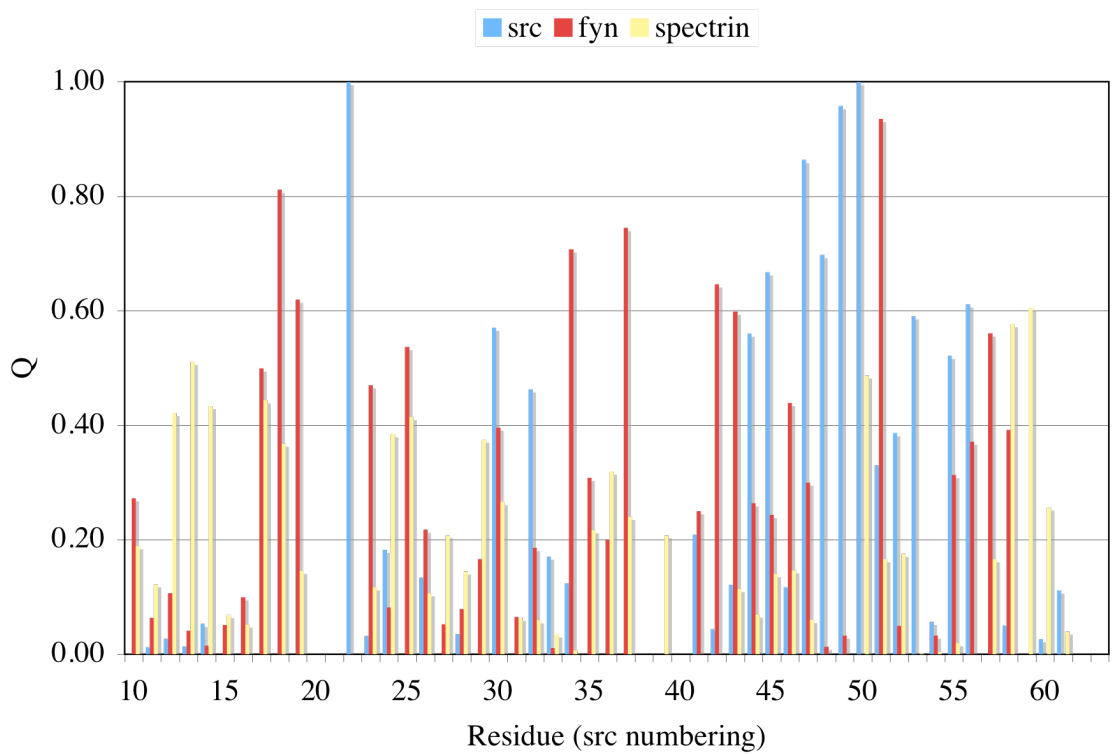

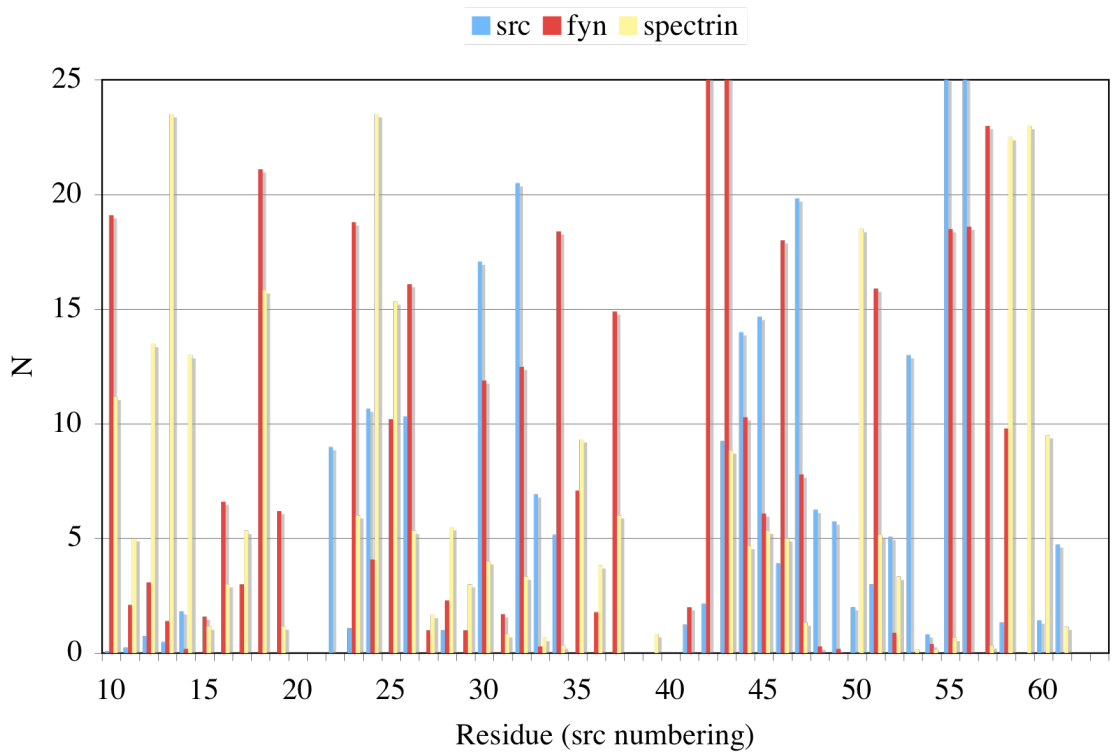

Supplemental Figure 2. Correlation (R=0.85) between Go and hybrid potential $p_{fold}$ values for 100 structures (some data points overlap).

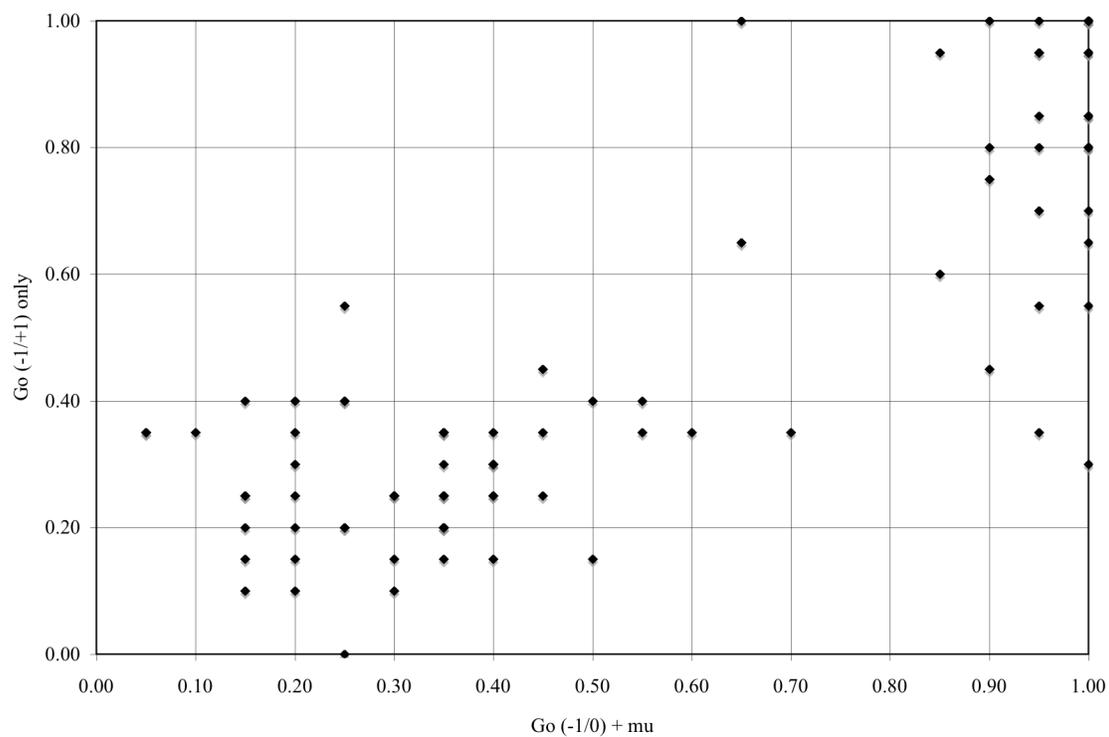

Supplemental Table 1. 84 atom types.

| atom | residue | type | atom | residue | type |
|---|---|---|---|---|---|
| CB | ALA | 0 | CB | MET | 43 |
| CB | ARG | 1 | CG | MET | 44 |
| CG | ARG | 2 | SD | MET | 45 |
| CD | ARG | 3 | CE | MET | 46 |
| NE | ARG | 4 | CB | PHE | 47 |
| CZ | ARG | 5 | CG | PHE | 48 |
| NH1 | ARG | 6 | CD1 | PHE | 49 |
| NH2 | ARG | 6 | CD2 | PHE | 49 |
| CB | ASN | 7 | CE1 | PHE | 50 |
| CG | ASN | 8 | CE2 | PHE | 50 |
| OD1 | ASN | 9 | CZ | PHE | 51 |
| ND2 | ASN | 10 | CB | PRO | 52 |
| CB | ASP | 11 | CG | PRO | 53 |
| CG | ASP | 12 | CD | PRO | 54 |
| OD1 | ASP | 13 | CB | SER | 55 |
| OD2 | ASP | 13 | OG | SER | 56 |
| CB | CYS | 14 | CB | THR | 57 |
| SG | CYS | 15 | OG1 | THR | 58 |
| CB | GLN | 16 | CG2 | THR | 59 |
| CG | GLN | 17 | CB | TRP | 60 |
| CD | GLN | 18 | CG | TRP | 61 |
| OE1 | GLN | 19 | CD1 | TRP | 62 |
| NE2 | GLN | 20 | CD2 | TRP | 63 |
| CB | GLU | 21 | NE1 | TRP | 64 |
| CG | GLU | 22 | CE2 | TRP | 65 |
| CD | GLU | 23 | CE3 | TRP | 66 |
| OE1 | GLU | 24 | CZ2 | TRP | 67 |
| OE2 | GLU | 24 | CZ3 | TRP | 68 |
| CB | HIS | 25 | CH2 | TRP | 69 |
| CG | HIS | 26 | CB | TYR | 70 |
| ND1 | HIS | 27 | CG | TYR | 71 |
| CD2 | HIS | 28 | CD1 | TYR | 72 |
| CE1 | HIS | 29 | CD2 | TYR | 72 |
| NE2 | HIS | 30 | CE1 | TYR | 73 |
| CB | ILE | 31 | CE2 | TYR | 73 |
| CG1 | ILE | 32 | CZ | TYR | 74 |
| CG2 | ILE | 33 | OH | TYR | 75 |
| CD1 | ILE | 34 | CB | VAL | 76 |
| CB | LEU | 35 | CG1 | VAL | 77 |
| CG | LEU | 36 | CG2 | VAL | 77 |
| CD1 | LEU | 37 | CA | GLY | 78 |
| CD2 | LEU | 37 | N | XXX | 79 |
| CB | LYS | 38 | CA | XXX | 80 |

| CG | LYS | 39 | C   | XXX | 81 |
| CD | LYS | 40 | O   | XXX | 82 |
| CE | LYS | 41 | OXT | XXX | 83 |
| NZ | LYS | 42 | OCT | XXX | 83 |